\documentclass[12pt]{article}

\usepackage{graphics}

\begin{document}


\title{Persistence in a Simple Model for the Earth's Atmosphere Temperature Fluctuations} 

\author{ 
\bf Arturo Berrones\\
\small Posgrado en Ingenier\' \i a de Sistemas\\
\small Universidad Aut\'onoma de Nuevo Le\'on\\
\small AP 126 - F, Cd. Universitaria\\ 
\small San Nicol\'as de los Garza, NL 66450, M\'exico.} 

\maketitle



\begin{abstract}
  The effect caused by the presence of a number of distinct time scales in a simple
  stochastic model for the Earth's atmosphere temperature fluctuations is studied. 
  The model is described by a dissipative dynamics consisting of a set of coupled stochastic
  evolution equations. The system shows features that resemble recent observations. In
  contrast to other approaches, like autoregressive models, the fluctuations of the 
  atmosphere's
  temperature depend on parameters with clear physical meaning. 
  A reduced version of the model is constructed and its temporal 
  autocovariance function is explicitly written.
\end{abstract}

\section{Introduction}

Interaction among processes with several length and time scales is common to a variety of
complex systems. 
For instance, the long-range temporal correlations
found in signals from a variety of fields 
can be
associated with an interplay of a number of time scales 
~\cite{berrones,pyragas,hasselmann,diezemmann,berglund}. 
In particular,
it is an extended belief that the persistence
observed in the temperature fluctuations of the Earth's atmosphere is a consequence 
of its feedback with slower dynamical components in the climate system, like
the oceans and Earth's surface ~\cite{hasselmann}. 
Persistence at short time scales is related to the
everyday life observation
where similar weather conditions 
are likely to be experienced 
over a given region in a time scale of a few days. 
The existence of these short-term correlations make weather forecasting
possible. 
The climate's
persistence is also found for larger time scales,
however its characterization is a more difficult task ~\cite{kurnaz}.
In some recent experiments, the temperature records from different 
places around the globe 
have been analyzed. The observations indicate the existence of universal 
power-laws $C(t) \sim t^{- r}$
describing the correlations of the temperature fluctuations around its
mean seasonal value. 
Although there is some disagreement about the value of the
exponent $r$, it has been firmly established that the
persistence in temperature fluctuations can
indeed be characterized by power-law autocorrelation functions ~\cite{kurnaz}.
It has been reported by some authors that 
for time scales that range from $\sim 1$
to $\sim 25$ years,
correlations 
measured on data from meteorologic stations placed on small islands
decay with an exponent $r \sim 0.4$, 
while for continental stations data are closer to
$r \sim 0.7$ ~\cite{govindan}. 
According to other authors ~\cite{fraedrich}, the persistence is even 
more pronounced for the oceanic regions (being roughly characterized by a
$1/f$ noise), while in the inner continents $r \sim 1$ 
(in terms of the power spectrum, a white noise at low frequencies), with
a transition region in the coastal zones in which $r \sim 0.7$.

The emergence of long-range temporal correlations is a non-trivial feature
that can be used to test models of the Earth's climate ~\cite{govindan}.
The understanding of the
long-range temporal correlations is 
fundamental because they characterize the interaction
among the different climate components ~\cite{fraedrich}.
There is some controversy with respect to the description 
made by large scale models of the atmospheric temperature
variability ~\cite{fraedrich, vyushin, comment, reply, kurnaz}.
The purpose of the present
Letter is to introduce a conceptual stochastic model for the 
fluctuations of the
Earth's atmospheric
temperature, or more precisely, its radiated energy
(for a survey on conceptual climate models see, for instance, 
Imkeller and Monahan ~\cite{imkeller}). 
It will be shown that the model 
displays qualitative features that closely resemble observations. 
In contrast
with other simple stochastic models of atmospheric temperature fluctuations, 
like autoregressive models ~\cite{kiraly, caballero}, all the parameters
of the presented model have direct physical interpretation. 
Additionally, as it will be discussed below,
the model introduced here also reveals statistical
features that are reminiscent of recent observations on the spatial
structure of the climate network.  
Therefore, this
work is intended to be a contribution towards the construction of realistic and
unexpensive algorithms for Earth's climate simulation. 

The model is based on energy balance ~\cite{cushman}. On Earth, as in other planets
with a solid crust, the influx of solar radiation is 
balanced by the outflow from the surface and the atmosphere. In the simplest description
of this process,
the Earth is treated as a single point. Let us denote by $y$ and $x$ the global
averages of the radiation emitted by the atmosphere and by the surface 
(oceans and land),
respectively. A fraction of the Sun's total radiation is 
immediately reflected back into space
and another is absorbed by the
atmosphere. The remainder of the flux is transmitted through the atmosphere
and reaches the surface, which in turn absorbs some of the radiation
and reflects the rest. The radiation absorbed by the surface is then radiated back
as heat. The surface radiates the absorbed energy in the infrared (IR) region of
the spectra. It turns out that the atmosphere is not transparent to IR radiation,
essentialy due to the presence of the so-called greenhouse gases. Let us denote by
$a$ the fraction of IR radiation absorbed by the atmosphere. 
A fraction $b$ of the
total radiation absorbed by the atmosphere is directed towards the
surface and the rest is finally lost into space.
All these considerations
are put together in the well known zero dimensional energy balance model:
\begin{eqnarray}\label{zero}
y=A+ax, \\ \nonumber
x=B+by,
\end{eqnarray}

\noindent
where constants $A$ and $B$ are the net 
contribution made to $y$ and $x$ by the
solar radiation flux, taking into account that some
heat is removed by water evaporation from the surface. 
The constant $a$ is called the
infrared (IR) absorption coefficient.
All the constants in Eqs. (\ref{zero}) are calculated by
averaging over a year and over the entire Earth's surface. Assuming a blackbody
process, the average atmosphere's temperature is given by $y = \nu {\it T}^{4}$,
where $\nu $ is the Stefan--Boltzmann constant. 
In spite of its simplicity, the 
zero dimensional energy balance model is capable of predicting with very good
accuracy the mean Earth's surface temperature.
Another interesting prediction of the 
zero dimensional energy balance model is the increment of the mean
temperature as the coupling parameters $a$ and $b$ grow.

In this Letter a spatially extended and time-dependent 
generalization of model (\ref{zero}) is introduced. The Letter is organized as 
follows: in Sec. \ref{model} the model is introduced and the temporal and
spatial correlations
are discussed numerically. A reduced version of the model is constructed and
formally solved in the framework of the Langevin
approach. A discussion of the statistical properties of the solution
is given. 
Some conclusions and future directions are discussed in 
Sec. \ref{conclusions}.

\section{The Model} \label{model}

Model (\ref{zero}) is generalized by the assumption that energy balance is 
satisfied locally and a transient time is necessary in order to achieve
a stationary state. 
A set of $N$ cells is considered. In each cell, atmosphere interacts with the surface through
the local atmosphere's IR absorption coefficient $a_n$ and the local
fraction of heat that 
the atmosphere returns towards the surface, $b_n$. Each component, atmosphere and surface,
has its own intrinsic local response time. The cells of each component interchange radiation
via a diffusive process. The model is written as
\begin{eqnarray} \label{model}
\dot{y}_n = d_1\Delta y_n - \lambda _n [y_n - (A_n + a_n x_n) ] + \varepsilon _n (t),
\\ \nonumber
\dot{x}_n = d_2\Delta x_n - \gamma _n [x_n - (B_n + b_n y_n)].
\end{eqnarray}

\noindent
In this equation $(\lambda _n)^{-1}$ and $(\gamma _n)^{-1}$ are the local response times of the 
atmosphere and the surface, respectively. The symbol $\Delta$ is the discrete
Laplacian and $d_1$, $d_2$ represent the diffusion coefficients of
each component. The term $\varepsilon _n (t)$ is a Gaussian white noise, without
correlations between different cells. The meaning of the rest of the terms follow
from the zero dimensional energy balance model (\ref{zero}). In particular,
$y_n$ gives
the radiation emitted by the atmosphere in the site $n$ at a given time. 
The atmosphere is expected
to have shorter intrinsic response times than that of the surface. 
The radiation emitted by the surface at time $t$ in cell $n$ is
represented by $x_n$. The noise reflects the more
rapid variations, or $weather$. Periodic boundary conditions are taken.
The constants $\lambda _n$, $\gamma _n$, $A_n$, $B_n$, $a_n$ and $b_n$ are
assumed to be independent variables, such that averages over index (in the
limit $N \to \infty$) 
give
the corresponding values of the 
parameters of the zero dimensional energy balance model, $A$, $B$, $a$, $b$; 
and the effective inverse response times for the atmosphere and the surface. The first 
important thing to notice with model (\ref{model}) is that it recovers the zero dimensional
energy balance model. This can be seen by averaging Eq. (\ref{model}) over the cells in the
limit $N \to \infty$.  The following reduced version of Eq. (\ref{model}) is obtained:
\begin{eqnarray}\label{motion}
\dot{y}=-\lambda [y-(A+ax)]+\varepsilon(t), \\ \nonumber
\dot{x}=-\gamma [x-(B+by)].
\end{eqnarray}

\noindent
From the fact that 
Eq. (\ref{motion}) represents an overdamped dynamics in a parabolic potential, the
system converges to a stationary state with mean value given by the solution of
the zero dimensional energy balance model. This result shows that model (\ref{model})
is capable of representing with good approximation the mean behavior of energy in
the coupled
atmosphere -- surface system. Now it will be argued that Eq. (\ref{model}) can
give realistic descriptions of temperature fluctuations as well. A large variability
over the intrinsic characteristic times of the surface 
around the globe is expected to exist,
as a consequence of the different response times present in the geosphere,
hydrosphere, cryosphere and biosphere.  
At first instance, this situation is 
modeled by treating the $\gamma _n$'s 
like independent random variables taken from a uniform probability 
distribution.
On the other hand, under the basis of the relative homogeneity of the atmosphere
composition, it will be assumed by now that the response time to perturbations
of the atmosphere is the same in all cells.
In what follows, the value $\lambda_n = \lambda = 1$ is used.
Under these assumptions, the time unit is defined as the mean 
atmospheric response time. 
The model represents the interaction between a hierarchy of time scales,
ranging from minutes to days to geological times. From this
point of view, it would be therefore reasonable to assume that $1/ \lambda$
lies in the intermediate scales, roughly in the range from
weeks to months. This and other important aspects about the definition of
the parameters in the model are intended to be refined by the author
in the near future by close cooperation with climate experts. At this point the
main goal is limited to explore the capabilities of the model (\ref{model}), in order
to give qualitatively realistic statistical descriptions of the temperature 
fluctuations present in the Earth's atmosphere.

\begin{figure}[htbp] 
\vskip0.5cm
\centering{\resizebox{6cm}{!}{\includegraphics{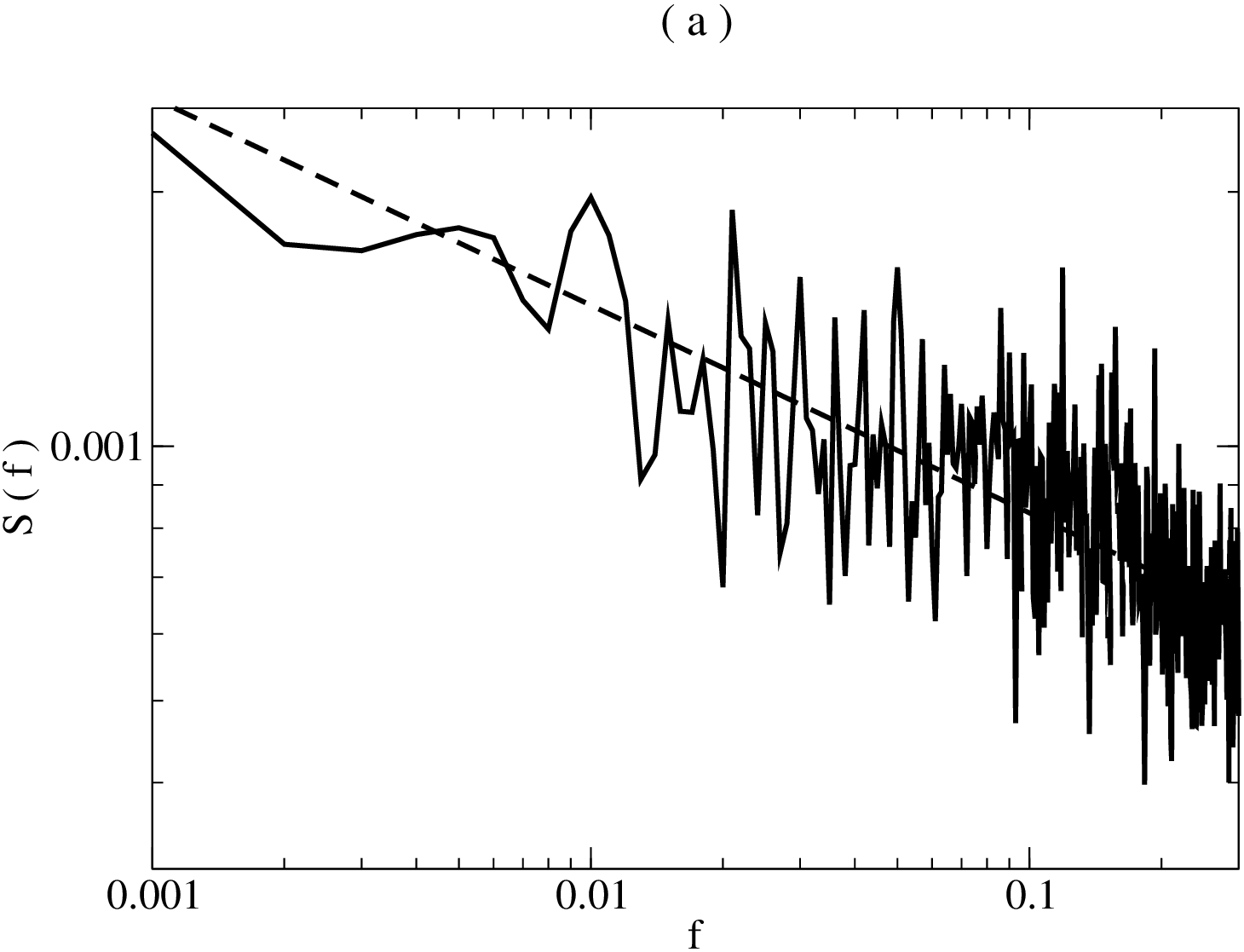}}}
\hskip0.5cm
\centering{\resizebox{6cm}{!}{\includegraphics{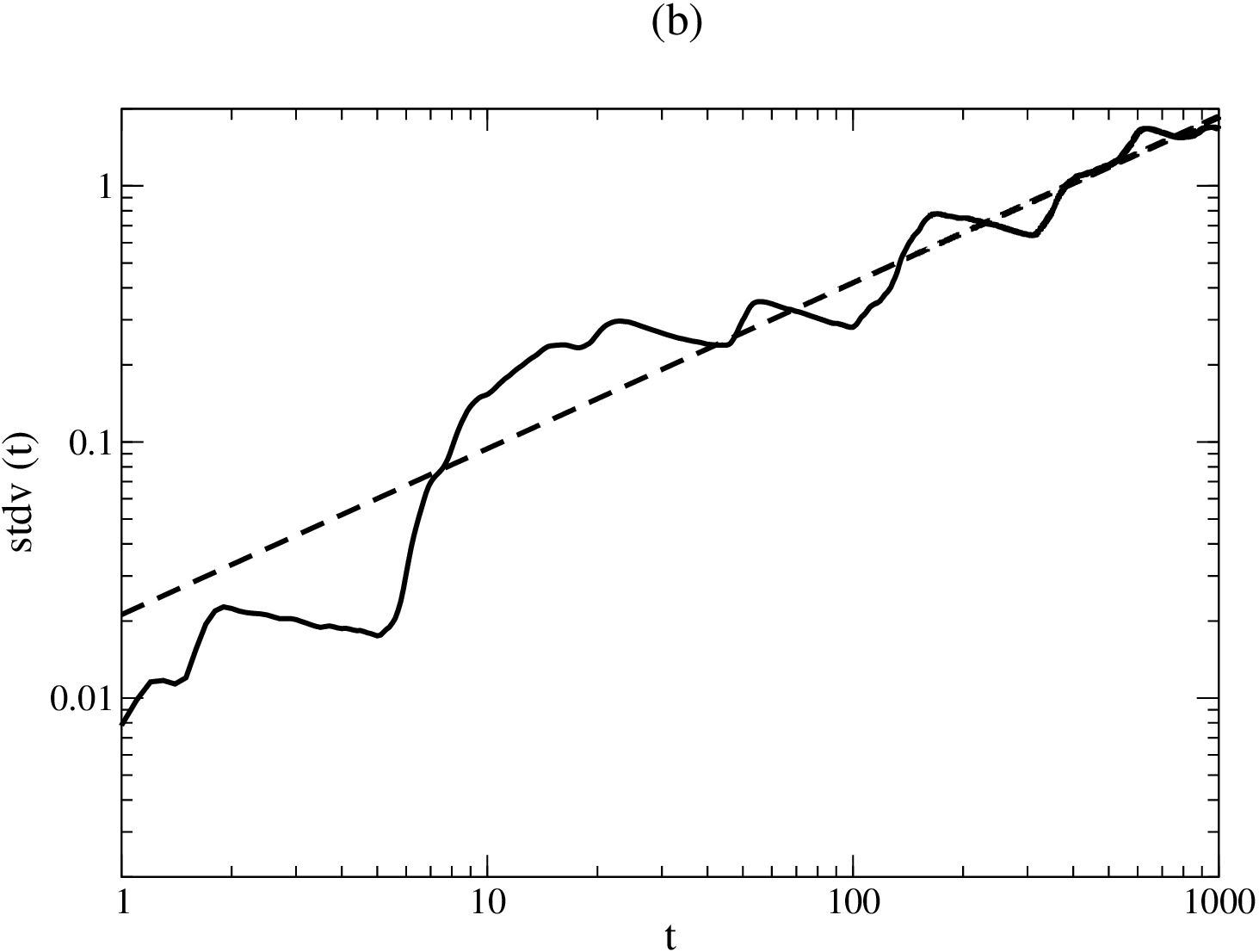}}}
\caption{\label{fig1} (a): Log--log plot of the 
power spectrum of a temperature signal generated by the term $y_3(t)$ of
model (\ref{model}). The parameters are as discussed in the text. The 
power spectrum is consistent with the power-law $S(f) \sim f^{-0.25}$. This indicates
a power-law decay of the autocorrelation function given by $C(t) \sim t^{-0.75}$ for
time scales between five and $1000$ time units.
(b): An alternative way to estimate the autocorrelation function for the same 
situation as in (a). The standard deviation as a function of
time of the sum of the temperature signal is plotted in a log--log graph. 
The graph shows a clear
difference with respect to the behavior expected from a random walk. The standard 
deviation is consistent with $\sigma (t) \sim t^{-0.65}$, which indicates
$C(t) \sim t^{-0.7}$ for time scales up to $1000$ time units.} 
\end{figure}

Figure \ref{fig1}(a)
is a log--log plot of the power spectrum of the time series of the atmosphere's temperature in a 
particular cell. This time series is obtained from the corresponding time series for
$y_n (t)$ under the assumption of a blackbody process. 
The parameters $a_n$, $b_n$, $A_n$ and $B_n$ are left constant and set to
their experimental global averages: $a_n = a = 0.96$, $b_n = b = 0.61$,
$A_n = A = 179.36$ ${\rm W}/{\rm m}^{2}$ and $B_n = B = 47.82$ ${\rm W}/{\rm m}^{2}$ ~\cite{cushman}.
The values $\gamma _n$ are drawn from a uniform distribution on the
range $(0, 0.2)$. The noise values $\varepsilon _n (t)$ are uniformly
distributed over the interval $(-1, 1) W/m^{2}$. 
The diffusion coeffiecients are taken as $d_1 = d_2 = 1$.
System size is $N=50$.
The power spectrum is consistent with 
a power law at low frequencies, with exponent $\sim -0.25$. This exponent of the 
power spectrum implies a power-law decay of correlations at large times, 
$C(t) \propto t^{- r}$, with $r \sim 0.75$. 
An alternative approach is given in Fig. \ref{fig1}(b). In order to characterize
the correlations, the sum of the temperature values on time at a given cell is
studied, resulting in:
\begin{eqnarray}\label{rw}
Y_n(t)=\sum_{\tau = 1}^{t} T_n(\tau),
\end{eqnarray}

\noindent 
where $T_n(\tau)$ stands for the atmosphere's temperature at cell $n$ and time $\tau$. 
The signal $Y_n(t)$ is then compared to a random walk. In Fig. \ref{fig1}(b)
the standard deviation of $Y_n(t)$ as a function of time is plotted.
For large times (roughly greater than $50$ time units), $\sigma \sim t^{0.648}$, 
which implies a power-law decay of 
the autocorrelation function at large times, with exponent
$\sim 0.7$. 
This result is consistent with Fig. \ref{fig1}(a).

An analysis of the radiation spectrum has been carried out
for the same experimental setup as above.
Results indicate that radiation and temperature spectra
are basically equivalent, differing at most by a 
normalization factor.
This numerical finding is consistent with recent studies that suggest
that a strongly correlated 
signal preserves its correlation properties after an even polynomial
transformation ~\cite{stanley}. 

\begin{figure}[htbp] 
\vskip0.5cm
\centering{\resizebox{6cm}{!}{\includegraphics{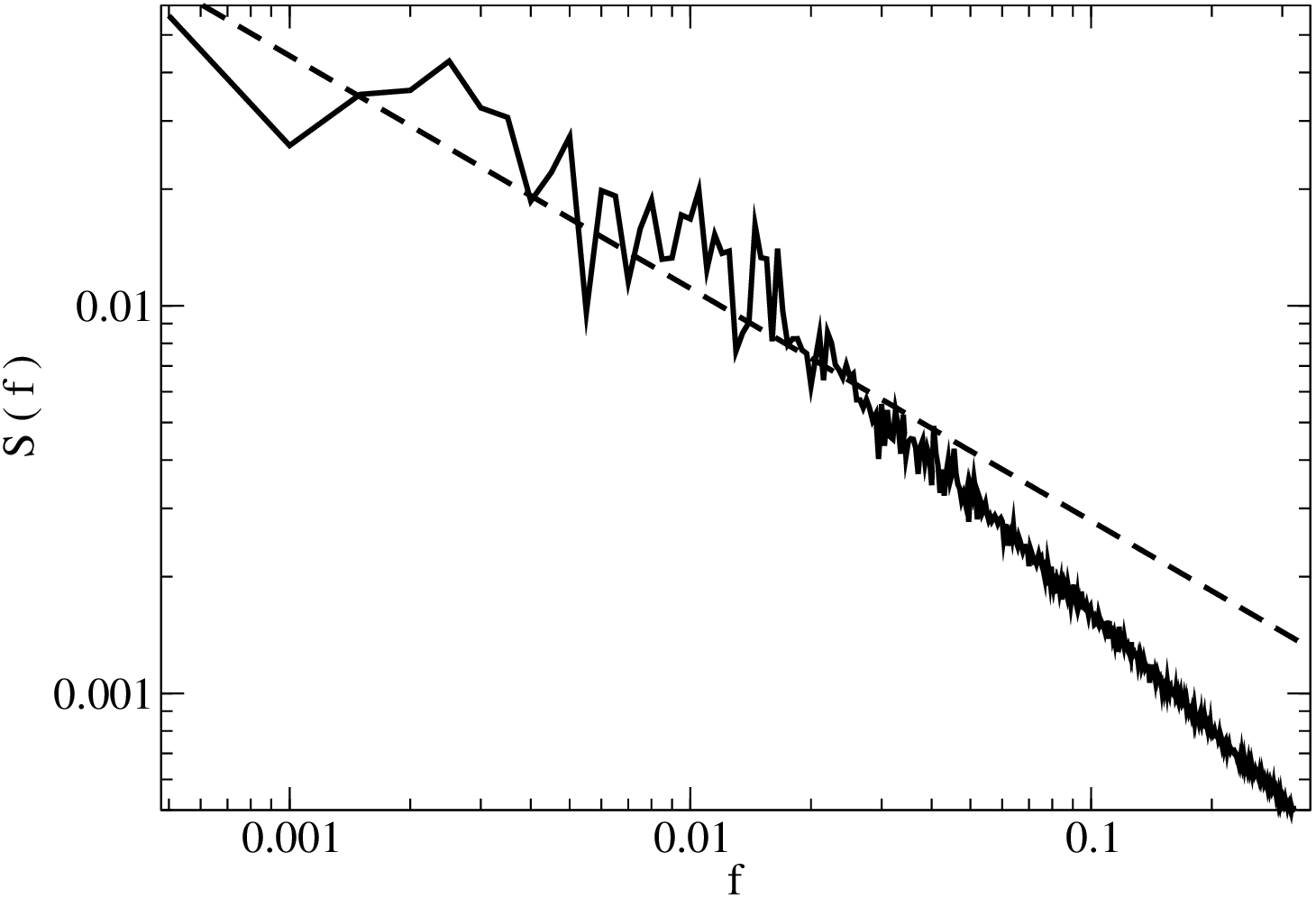}}}
\caption{\label{fig1c} Log--log plot of the 
power spectrum of a temperature signal generated by the term $x_3(t)$ of
model (\ref{model}). The parameters are the same as in Fig. 1, 
except for $d_2=0$. 
A power-law $S(f) \propto f^{-0.6}$ is plotted for comparision.} 
\end{figure}

The model is also capable of 
showing scaling behavior for surface temperature.
As an example, the power spectrum of surface temperature fluctuations 
of an induvidual cell is shown
in Fig. \ref{fig1c}. The system has the same parameter values
as before, but $d_2=0$. Notice that with this choice of parameters, the  
coupling with the atmosphere is essential for the emergence of
scaling in the surface temperature.

\begin{figure}[htbp] 
\vskip0.5cm
\centering{\resizebox{6cm}{!}{\includegraphics{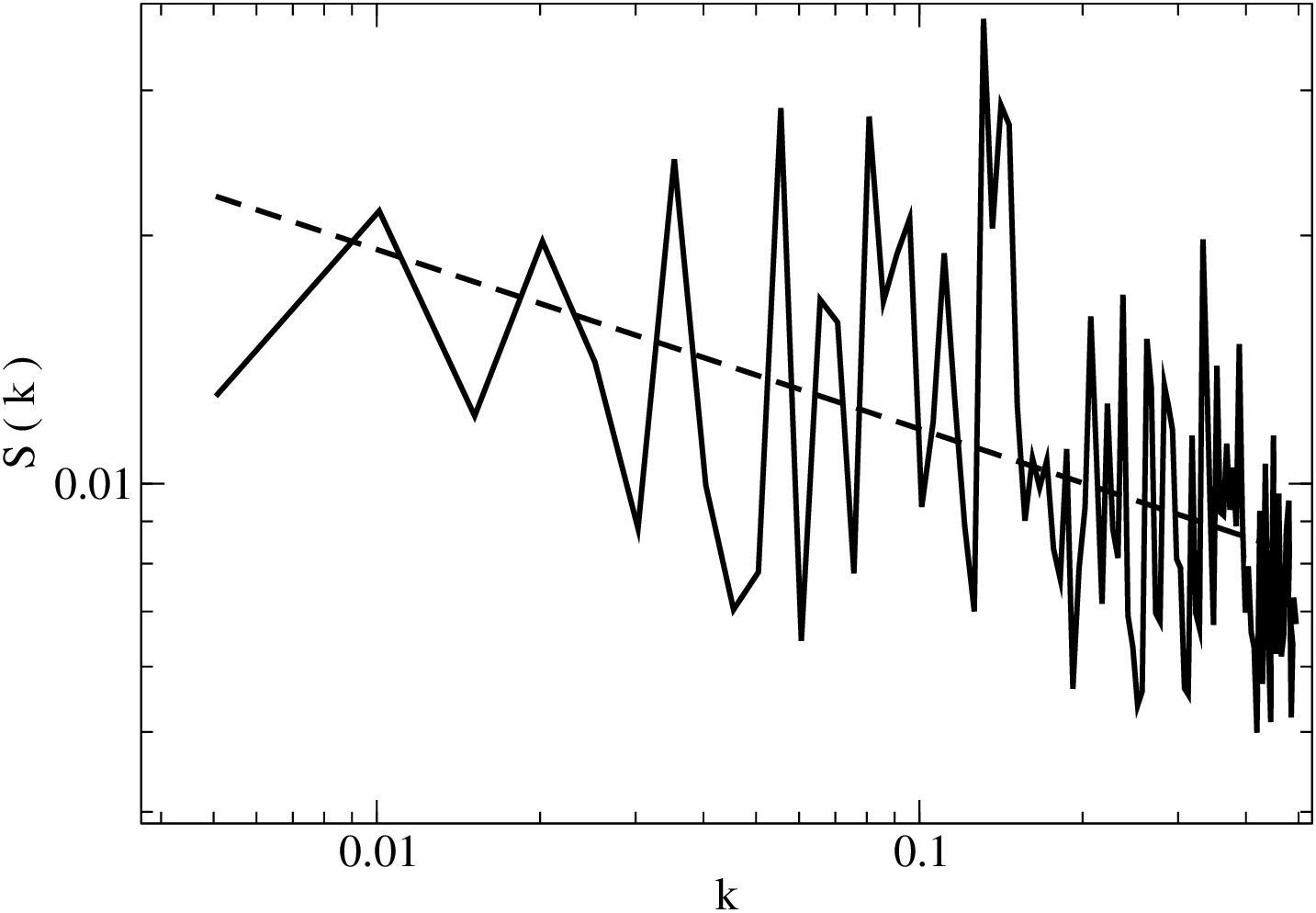}}}
\caption{\label{fig2} Log--log plot of the power spectrum of the spatial 
vector of temperature values vector at a fixed time. The parameters are as
discussed in the text. For this situation, in which there is no local
variability in the parameters (besides $\gamma _n$), the power spectrum
indicates a power-law decay of the spatial autocorrelation
function for scales in the range from one to $200$ cells, characterized 
by $C(n) \sim n^{-0.78}$.} 
\end{figure}

The model displays an interesting spatial structure. 
In Fig. \ref{fig2} a case is considered in which the system size is 
$N=200$ and the other parameters are the same as in the case presented
in Fig. \ref{fig1}. 
The spatial autocorrelation
function $C(n)$
is inferred from the power spectrum of the 
temperature values vector at a fixed time.
A least-square fit of the power spectrum to a power-law function
indicates that $S(k) \sim k^{-0.22}$, which implies
that the spatial autocorrelation function
can be roughly characterized by $C(n) \sim n^{-0.78}$
for spatial scales in the range from one to $200$ cells.
Further numerical analysis
of the same model setup
indicates that if larger spatial scales are considered,
the power spectrum displays a crossover to 
white noise
at low frequencies.

\begin{figure}[htbp] 
\vskip0.5cm
\centering{\resizebox{6cm}{!}{\includegraphics{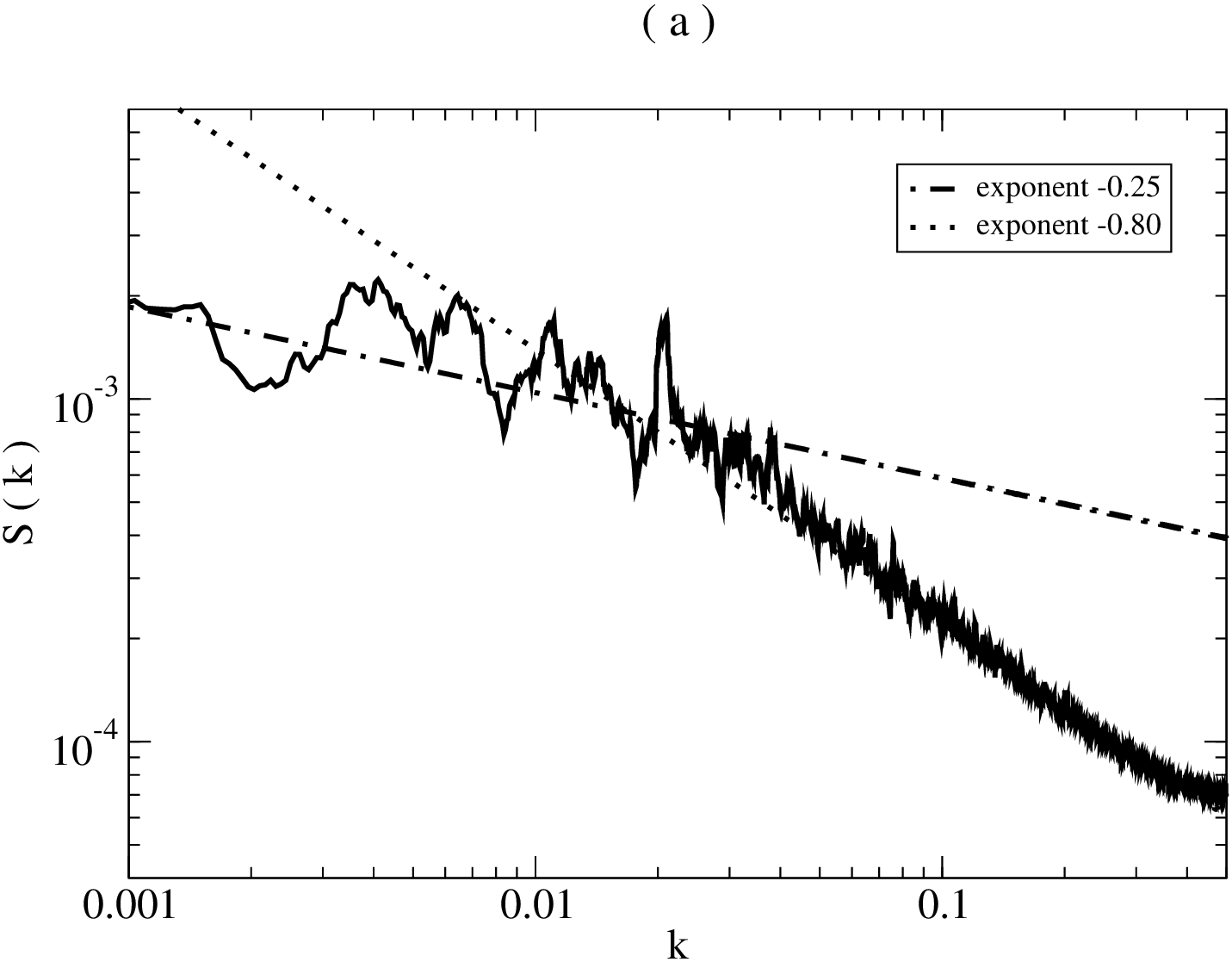}}}
\hskip0.5cm
\centering{\resizebox{6cm}{!}{\includegraphics{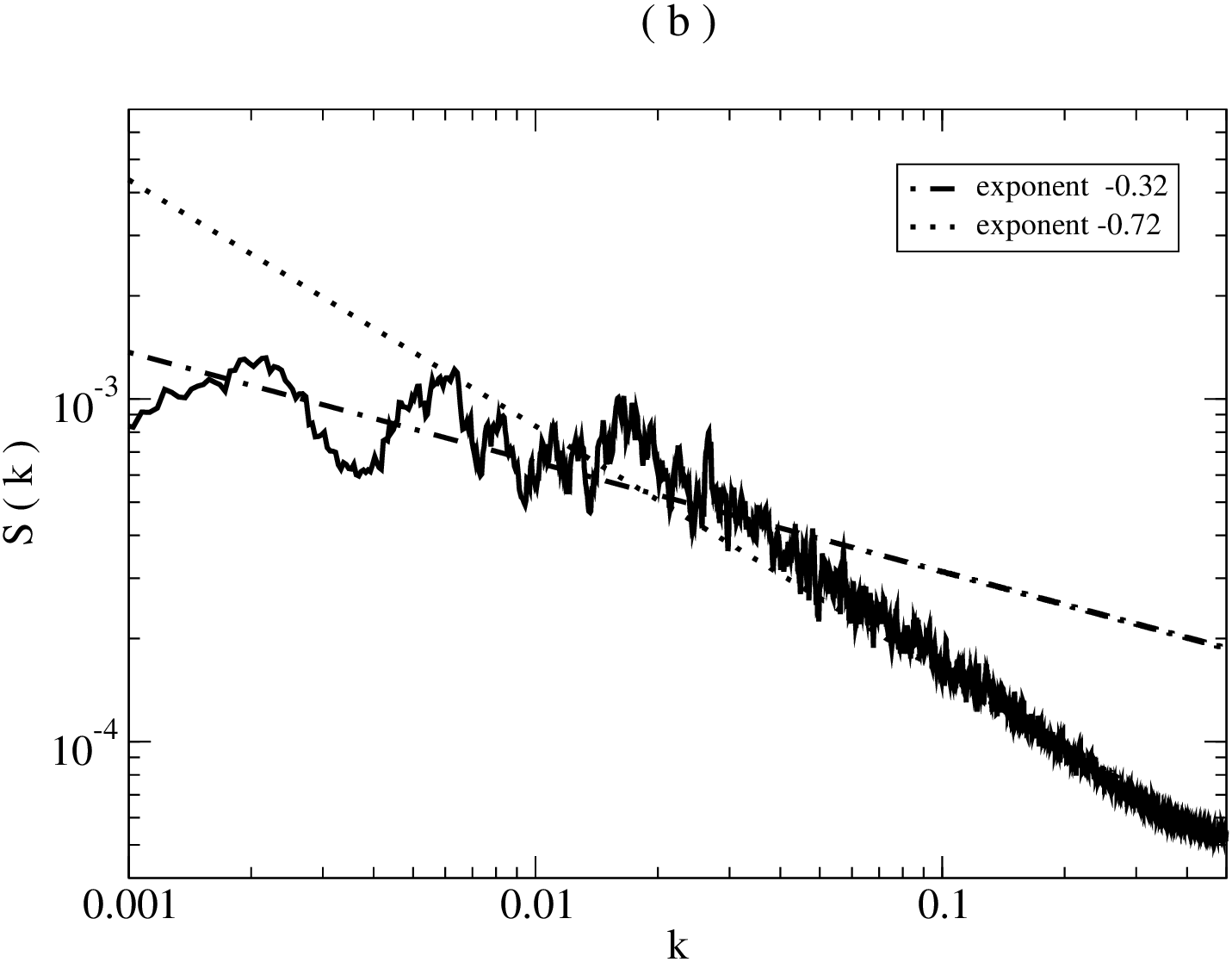}}}
\caption{\label{fig3} This situation is similar to Fig. \ref{fig2}, except
from the fact that some parameters besides the $\gamma _n$'s have local variability.
In (b)
the local variability and the mean coupling parameter $a$ 
have larger values
than in (a).} 
\end{figure}

In a more realistic description, the parameters (besides $\gamma _n$)
must have some local variability due, for instance, to differences in the
Earth's albedo and in the solar radiation flux over different regions.
As an example, a case in which $A_n = A + \epsilon _n$ and 
$B_n = B + \epsilon ^{'}_{n}$ is discussed in Fig. \ref{fig3}(a).
The $\epsilon$'s are independent random variables uniformly 
distributed in the range $(-1, 1)$ ${\rm W}/{\rm m}^{2}$. The other parameters are chosen
as before. The power spectrum presented in Fig. \ref{fig3}(a)
indicates a crossover between two different scaling regimes.
For scales from $50$ to $1000$ cells the correlation function
is consistent with $C(n) \sim n^{-0.75}$, while for shorter
scales $C(n) \sim n^{-0.2}$.
A situation in which the
coupling parameters have also local variability
is presented in
Fig. \ref{fig3}(b). The values of $b_n$ are set as
$b_n = b + \rho _{n} $, where $\rho _{n}$ is uniformly distributed
over the interval $(-0.15, 0.15)$. 
The parameters $a_n$ are taken as $a_n = 0.98$ for all $n$, which
implies a stronger mean coupling.
The other parameters are the same
as in Fig. \ref{fig3}(a). In Fig. \ref{fig3}(b) the correlation
decay is faster for the short scales and slower for the large scales in
comparision to Fig. \ref{fig3}(a). This effect can be interpreted as an 
increment of large scale coherence as the mean coupling
grows, while the spatial coherence 
at short scales decreases due to the increment in
the local variability. 
The scaling of 
the spatial autocorrelation 
function displayed by the model 
is reminiscent of recent observations on the spatial
structure of the climate network, which indicate that nodes in the climate
system conform to a network with the {\it small-world} property  ~\cite{tsonis}. 
This property
is related to the presence of significant correlations between distant nodes.
 
The numerical findings strongly suggest 
that model (\ref{model}) shows qualitative features that are close
to the observations. However, 
a more
precise definition of the parameters is needed. For instance, a different choice
of the scale separation between the $\gamma$'s and $\lambda$ in general
lead to different properties of the autocorrelations.  
Another aspect to be refined is concerned with the 
already mentioned spatial variability of the parameters. This
question is closely related to the definition of the 
size associated
to cells. 
In a realistic model setup, the parameter values 
come from
spatial averages over the region $n$. 
In the simple $1d+1$ situation discussed here, those would be 
global averages over a given
latitudinal interval.
As already mentioned,
these and other relevant questions are intended to be
investigated by the author in the near future, working in close contact with
climate experts.

In order to gain insight into model (\ref{model}) it is discussed analytically its 
reduced version given by Eq. (\ref{motion}).
Without loss of generality, the constants $A$ and $B$ are chosen equal to zero. 
The term
$\varepsilon(t)$ is a Gaussian white noise,
defined through the moments
$\left < \varepsilon(t) \right > = 0$, 
$\left < \varepsilon(t) \varepsilon(t^{'}) \right > = D \delta (t-t^{'})$ 
and with all higher moments equal to zero. The diffusion constant
is a parameter that measures the strength of the noise. The function
$\delta (t-t^{'})$ is a Dirac's delta.
In the absence of coupling $x$ and $y$
simply
converge exponentially to the stationary state $\left< y \right>=x=0$, 
$\left< y^{2} \right>= \frac{D}{2\lambda}$
with characteristic  
times $\frac{1}{\lambda}$ and $\frac{1}{\gamma}$. 
It is assumed that 
$\lambda > \gamma $ so one of the dynamics is $fast$ and the other is $slow$.
Strictly speaking, the system has three time scales, the third one being associated
with the noise. However this time scale has an infinite separation with respect to the other
two. 
In the language of control theory, we can view $y$ like an
output system with uncertainties (noise)
that has feedback with an input whose response time is different from the time scale of
$y$. 
Applying a Laplace transform
over time to Eq. (\ref{motion}) and assuming for 
simplicity the initial conditions $x(0)=y(0)=0$, one gets
\begin{eqnarray}\label{algebra}
s y(s)=-\lambda y(s) + \lambda a x(s) + \varepsilon(s), \\ \nonumber
s x(s)=-\gamma x(s) + \gamma b y(s).
\end{eqnarray}

\noindent Solving Eq. (\ref{algebra}) for $y(s)$ and $x(s)$,
and by the use of the Faltung theorem, 
a formal solution for $y(t)$ in terms of the noise is found.
This solution can be used to write explicitly the covariance function, that describes
the fluctuations around the mean value. 
\footnote{This covariance function can also be derived from a 
general initial value problem for a linear
system of stochastic differential equations
of arbitrary dimension.}
For large times, the covariance 
function is given by the
following expression:
\begin{eqnarray}\label{correlation}
\left < y(t)y(t+T) \right > =
q_1 e^{\mu _1 T} + q_2 e^{\mu _2 T},
\end{eqnarray}

\noindent where 
\begin{eqnarray}
q_1 =
\frac{D}{16K^2}(\lambda - \gamma + 2k)
\left (\frac{\lambda - \gamma + 2K}{\lambda + \gamma + 2K} + 
\frac{\gamma - \lambda + 2K}{\gamma + \lambda} \right ),
\end{eqnarray}

\begin{eqnarray}
q_2 =
\frac{D}{16K^2}(\gamma - \lambda + 2K)
\left (\frac{\gamma - \lambda + 2K}{\gamma + \lambda - 2K} + 
\frac{\lambda - \gamma + 2K}{\gamma + \lambda} \right ),
\end{eqnarray}

\begin{eqnarray}
\mu _1 = -\frac{\gamma + \lambda}{2} - K,
\end{eqnarray}

\begin{eqnarray}
\mu _2 = -\frac{\gamma + \lambda}{2} + K,
\end{eqnarray}

\begin{eqnarray}
K = \sqrt{\left ( \frac{\gamma +\lambda}{2}\right )^{2} - \gamma \lambda (1-ab)}.
\end{eqnarray}

\noindent In the region of interest of the parameter space, correlations 
decay monotonically with a 
characteristic time
\begin{eqnarray} 
\tau =
\frac{1}{ \left( \frac{\gamma + \lambda}{2} \right)
-
\sqrt{\left(\frac{\gamma + \lambda}{2} \right)^{2}
- \gamma \lambda (1-ab) }  }.
\end{eqnarray}

\noindent  It must be realized that $\tau$ is greater
than any of the two intrinsic times. Therefore, because
of the feedback there is an emergence of large memory. 
The characteristic time becomes infinitely large when 
$\gamma \to 0$ with $\lambda$ finite. 
This result is consistent with
previous works on reduced models of temperature fluctuations ~\cite{fraedrich2}.   

\begin{figure}[htbp] 
\vskip0.5cm
\centering{\resizebox{6cm}{!}{\includegraphics{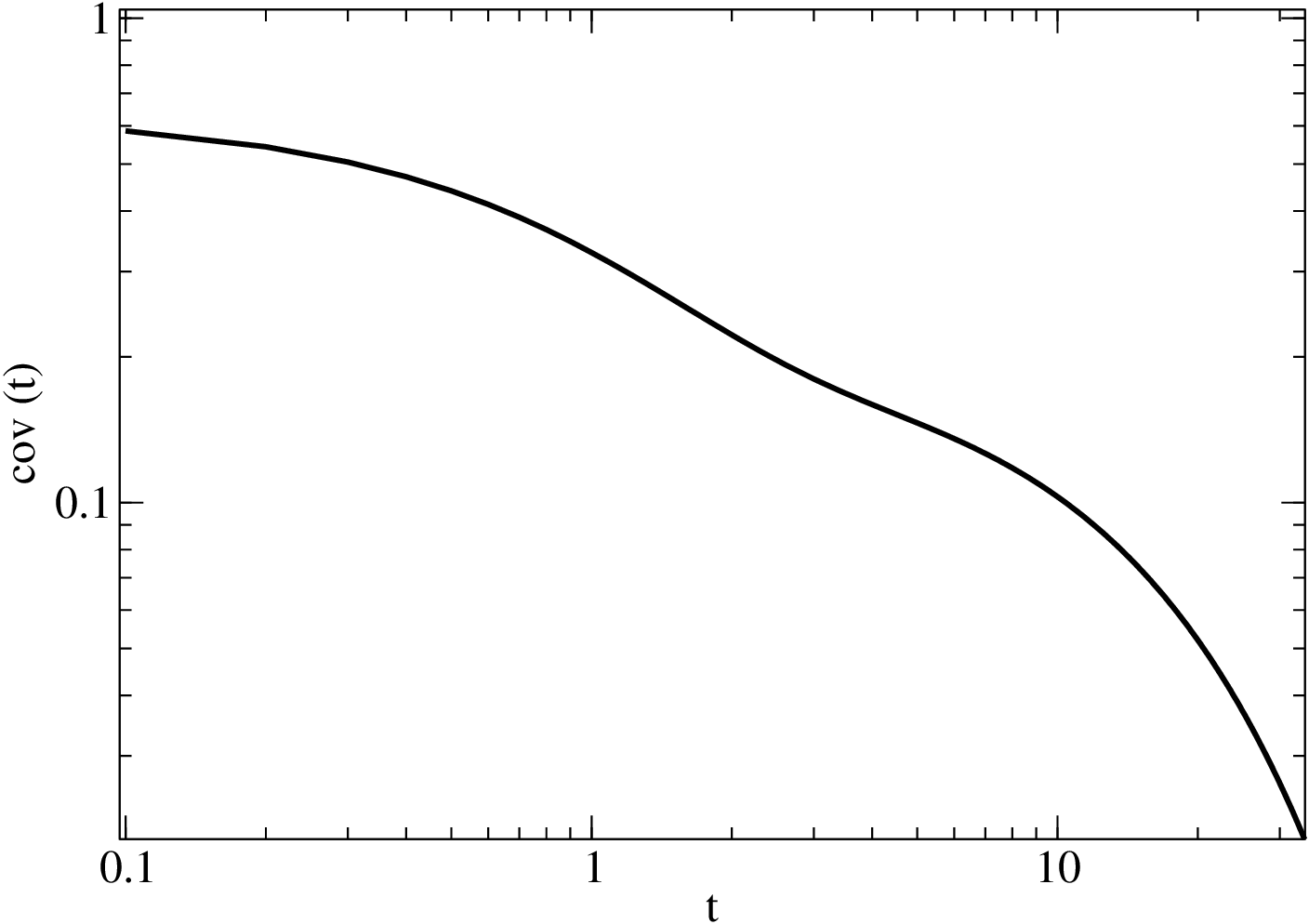}}}
\caption{\label{fig4} Log--log plot of the covariance function
Eq. (\ref{correlation}) with parameters as discussed in the
text. The covariance cannot be fitted to a single
exponential for a time interval greater than any of the two 
intrinsic times.} 
\end{figure}

The reduced model typically shows a region in which none of the two 
intrinsic time scales
is dominant and correlations cannot be adequately fitted by a single
exponential. For instance, with
$\gamma = 0.2$, $\lambda = 1$, $a = 0.96$, $b = 0.64$ and $D = 1 (\frac{W}{m^{2}})^{2}$,
the covariance function is not exponential for time scales
approximately an order of magnitude greater than the intrinsic time
of the fast variable, as Fig. \ref{fig4} shows. 
Approximation to a power-law or other
types of slow decay by a sum of exponentials with
different characteristic times has been discussed 
in several fields ~\cite{anderson,bouchaud}. 
In particular, this mechanism has been already proposed in 
~\cite{caballero}, in order to explain the persistence found in the atmosphere's
temperature record, by fitting the coefficients of
a 3d AR(1) type process to data.

\section{Conclusions}\label{conclusions}

The features shown by
the spatially extended stochastic model
presented here motivates
the construction of realistic and simple algorithms for the 
prediction of the Earth's temperature distribution and fluctuations.
In this spatially extended situation the parameters
vary locally, so there is a 
number of characteristic times. 
In order to gain insight on the extended model, a reduced version of it 
has been
constructed and the covariance function explicitly written.

One of the future directions of the work is to conduct
a more general study of the spatially extended model, in close connection to climate
research to have plausible parameter values. A study of the presented model in the context
of general systems with several time scales is also intended.
The study of such systems is important in fields like control theory,
inhomogeneous media and predator--pray systems among others
~\cite{berrones,pyragas,hasselmann,diezemmann,berglund}.
   
\section*{Acknowledgments}
The author is grateful for
the valuable comments given by the unknown reviewers of the present Letter.
The author acknowledges partial financial support by SEP
under project PROMEP/103.5/05/372, CONACYT under 
project J45702-A and UANL-PAICYT.

\end{document}